# Topological quantum catalyst: the case of two-dimensional traversing nodal line states associated with high catalytic performance for hydrogen evolution reaction

Lirong Wang,[1,2] Xiaoming Zhang[1,2*], Weizhen Meng,[1,2] Ying Liu,[2] Xuefang Dai,[2] and Guodong Liu[1,2*]

[1]State Key Laboratory of Reliability and Intelligence of Electrical Equipment, Hebei University of Technology, Tianjin 300130, China

[2]School of Materials Science and Engineering, Hebei University of Technology, Tianjin 300130, China

Correspondence: zhangxiaoming87@hebut.edu.cn; gdliu1978@126.com

**ABSTRACT:**

Topological quantum catalysts (TQCs), where metallic surface states from nontrivial band topology serve as the mechanism to favor heterogeneous catalysis processes, have been well demonstrated in three dimensional (3D) examples but have been rarely discussed in 2D scale. Here, we develop a design scheme to realize 2D TQCs with showing traversing nodal line at the Brillouin zone boundary, large Fermi arc on the edge, and nearly zero Gibbs free energy ($\Delta G_{H*}$) for hydrogen evolution reaction (HER). We demonstrate the 2D $Cu_2C_2N_4$ sheet is a such example. The material manifests an open nodal line traversing the whole $k$-path S-Y. It shows a long Fermi arc that spans the entire edge boundary, which is robust against spin-orbit coupling and the H adsorption. As the result, the edge of $Cu_2C_2N_4$ sheet is relatively active for HER catalysis with possessing a $\Delta G_{H*}$ as low as 0.10 eV, which is comparable with that of Pt and superior to other traditional catalysts and 3D TQCs as well. Our work offers an effective route to develop high performance HER catalysis without containing noble metals by utilizing 2D TQCs with traversing nodal line.

# 1. Introduction

Facing on the increasingly serious situation of global warming from fossil fuels, in recent years tremendous attention has been paid on innovating the energy carriers to reduce power-station emissions of greenhouse gases. Hydrogen is believed as an excellent renewable energy carrier candidate because it can offer high-energy density but immune from carbon emission [1-3]. For hydrogen production, decomposing water is a promising and environmentally friendly method. At the current, one of the most crucial research focuses for the hydrogen generation is to develop excellent electrocatalysts with high efficiency and good stability for hydrogen evolution reaction (HER) [4–8]. Precious metals especially Pt, have been proved to be high-efficiency HER catalysts [9]. The high efficiency can be traced back to their relatively low Gibbs free energy ($\Delta G_{H^*}$) during hydrogen adsorption, with locating nearly on the top of the HER volcano plot [10-13]. However, the main drawbacks of these catalysts lie in their scarcity and the associated high price, which highly limit their application in large-scale industrial hydrogen production. For such a consideration, exploring highly efficient HER catalysts without precious metals is urgently need.

Recently, known as topological quantum catalysts (TQCs), topological nontrivial materials are highly hoped for providing another reasonable catalytic mechanism for designing high-efficiency catalysts [14]. In TQCs, metallic surface states induced by nontrivial bulk band topology have been proved to be effective in favoring charge-transfer kinetics during catalytic process for hydrogen/oxygen evolution, CO oxidation, and so on [15,16]. Till now, various TQCs with different topological features have been proposed. As the representative examples, topological insulators $Bi_2Se_3$ and $Bi_2Te_3$ were reported effective for HER and CO catalysis [16,17]. In addition, Weyl semimetals TaAs, NdAs, and $Co_3Sn_2S$, multiplefold degenerate semimetals PtGa, PtAl and $Nb_2S_2C$, nodal line semimetal TiSi family were also proposed to be conducive to the HER process. Most of these proposals have been verified by experiments [18-25]. Especially, a recent work finds the topological state and catalytic properties in TQCs can be linked by the exchange current parameter $I_0$ [26]. One can notice that these TQC examples are all limited to bulk materials. Considering the facts that: i) similar with bulk materials, two-dimensional (2D) system can also offer nontrivial band topology and the associated metallic edge states [27,28], which may also the facilitate HER catalytic process; ii) comparing 3D catalysts, 2D ones have shown several advantages such as higher charge transfer efficiency because of the unique morphology in 2D scale in non-TQC process [29-31]. Then, one may naturally wonder that, can TQCs be extend to 2D scale?

Motivated by above discussions, in this work, we investigate the feasibility of developing 2D TQCs for HER catalysis. We first construct a design scheme, which includes different topological features in 2D scale and evaluates the corresponding nontrivial edge states and potential HER catalysts activity. Then we have successfully identified 2D $Cu_2C_2N_4$ sheet as a concrete example following the design scheme. The material takes a nodal line traversing through the whole Brillouin zone (BZ) boundary and features with long Fermi arc spanning the entire edge. Remarkably, our calculation show $Cu_2C_2N_4$ sheet indeed yields a relatively low $\Delta G_{H*}$ (0.10 eV), which is close to the value in noble metal Pt and situates near the top of the volcano plot for HER. These results reveal 2D TQC with traversing nodal line is a possible platform to develop HER catalysis without containing noble metals.

## 2. Computational methods

The numerical calculations in current work were performed within the Vienna ab initio Simulation Package (VASP) [32] in the framework of density-functional theory (DFT) [33]. The generalized gradient approximation (GGA) of the Perdew−Burke−Ernzerhof (PBE) method [34] was applied for the exchange–correlation potential. To avoid interactions among layers, we build a vacuum space larger than 20 Å for the crystal structure of $Cu_2C_2N_4$ sheet. The cutoff energy was adopted as 520 eV, and the BZ was sampled with Γ-centered $k$-point mesh of 11×11×1. The DFT-D2 method [34] was used to consider the long-range van der Waals interactions. The PHONOPY code [35] was used to calculate the phonon spectra of $Cu_2C_2N_4$ sheet. From the maximally localized Wannier functions [36,37] and the WannierTools package [38], topological features of edge states for $Cu_2C_2N_4$ sheet were calculated.

**3. Design scheme**

In 3D TQCs, the catalytic activity is found to be highly relative to the density of metallic surface states near the Fermi energy [16,17]. For example, chiral fermions with larger topological charge in PtAl can generate longer Fermi arc surface states and higher HER catalytic activity than traditional Weyl semimetals [39]. Further, Li *et. al.* argued nodal lines can in principle provide higher surface intensity than those by nodal points, which has induced a nearly zero $\Delta G_{H^*}$ in 3D nodal line TQC TiSi [25]. However, this theory may not be simply extended to 2D TQC.

In 3D system, topological band crossings can take the form of 0D nodal point, 1D nodal line, and 2D nodal surface [40-47]. In 2D system, only 0D and 1D band crossings exist [48-51]. For nodal points in 2D [see Fig. 1(a)], they produce a nearly zero electronic density in the plane. Due to the presence of Fermi arcs, they show a nonzero electronic density on the edge, where the edge electronic density is in principle relative to the length of the Fermi arcs. For traditional nodal line in 2D system [see Fig. 1(b)], although electronic density in the plane is enhanced, the edge electronic density is still not strong enough because of the partial Fermi arcs. Even worse, such closed nodal line is topologically trivial and sometimes does not carry definite edge states [52]. To capture the strongest edge electronic density, it requires the edge Fermi arcs transverse the whole BZ boundary. Such an occasion can be realized by open nodal lines [53]. For open nodal line in 2D system [see Fig. 1(c)], it can realize strong electronic density both in the plane and on the edge. Especially, the edge electronic density is unusually stronger than the cases in Fig. 1(a) and (b), because the Fermi arc in this occasion can transverse the entire edge. Therefore, 2D TQC with open nodal lines would be the best electrocatalyst for the HER, and should carry a $\Delta G_{H^*}$ close to zero.

**4. Results and discussions**

**4.1 Structure and stability of Cu$_2$C$_2$N$_4$ sheet**

Within this context, in the following we will show Cu$_2$C$_2$N$_4$ sheet is an ideal 2D TQC candidate with hosting open nodal lines, traversing edge Fermi arc states, and relatively low $\Delta G_{H*}$ for HER. The 2D Cu$_2$C$_2$N$_4$ sheet is firstly proposed by the 2D materials database (also known as 2DMatPedia), where Cu$_2$C$_2$N$_4$ monolayer along with nearly 2,000 monolayers have been demonstrated to be easily strippable from their 3D counterparts based on the geometry algorithm and exfoliation energy [54].

In Fig. 2(a), we show the crystal structure for Cu$_2$C$_2$N$_4$ sheet in the form of a 1×3 supercell. The shadowed region in Fig. 2(a) indicates the primitive cell of Cu$_2$C$_2$N$_4$ sheet. It obviously takes a rectangular lattice. From the symmetry view, the lattice structure belongs to the space group of *PMMA* (No. 51). One primitive cell of Cu$_2$C$_2$N$_4$ sheet contains two Cu atom, two C atom and four N atoms. From the top view, the bonding among Cu-N and C-N atoms forms two hexagonal lattice configurations. As shown by the side view, the hexagonal lattice configurations are fluctuant in the out-of- plane direction. The optimized lattice constants for Cu$_2$C$_2$N$_4$ sheet are $a$ = 9.40 Å, $b$ = 2.96 Å. The Cu-N and C-N bond length yields to be 1.99 Å and 1.24 Å, respectively.

Although 2DMatPedia has demonstrated Cu$_2$C$_2$N$_4$ sheet can be easily exfoliated from the bulk, it is also essential to verify whether the freestanding Cu$_2$C$_2$N$_4$ sheet is stable. For such consideration, we first estimate the thermal stability of Cu$_2$C$_2$N$_4$ sheet by using the *ab* initio molecular dynamic (AIMD) simulation. In this simulation, the 3×2 supercell of Cu$_2$C$_2$N$_4$ sheet is used with the temperature set as 300K. The AIMD simulation is totally carried out for 2000 fs with 0.5 fs as one step. As shown in Fig. 2(b), one can find that Cu$_2$C$_2$N$_4$ structure nicely retains during the simulation, suggesting at room temperature Cu$_2$C$_2$N$_4$ sheet carries an excellent thermal stability. In addition, the phonon spectrum for Cu$_2$C$_2$N$_4$ sheet is calculated to determine its dynamic stability, as displayed in Fig. 2(d). We can observe no virtual modes throughout the highly symmetric *k*-paths. This verifies that Cu$_2$C$_2$N$_4$ sheet is also dynamically stable.

**4.2 Topological band structure and edge states of $Cu_2C_2N_4$ sheet**

We show the electron band structure and the density of states (DOSs) of $Cu_2C_2N_4$ sheet in Fig. 3(a). We can find $Cu_2C_2N_4$ sheet has a metallic electronic structure with sizable DOSs at the Fermi level. From the orbital projected DOSs, we can find the conducting electronic states are mostly contributed by the Cu and N atoms. As shown in the band structure in Fig. 3(a), there are two bands locating near the Fermi level. The two bands overlap together along the $k$-path S-Y. For other $k$-paths, the two bands are well separated with each other. From the 3D plotting of the two bands [see Fig. 3(b)], we find they in fact form a nodal line. As shown in Fig. 3(c), the nodal line situates at the 2D BZ boundary and is characterized as an open nodal line. In Fig. 3(d), we show the edge states of the nodal line in $Cu_2C_2N_4$ sheet. We can find a long Fermi arc appears on the edge originating from the nodal line. To be noted, in this case, the Fermi arc is the longest for 2D nodal line, because it spans the whole edge.

After taking into account the spin-orbit coupling (SOC), we find the profile of the band structure does not have much change, as shown in Fig. 4(a). Except that, the doubly-degenerate bands in the S-Y path are slightly split. However, at the S and Y points the band degeneracy retains, which is required by symmetry. In Fig. 4(b), we show the enlarged band structure near the S and Y points. To be noted, in this occasion, the long Fermi arc on the edge still exists because of the presences of band crossing at the S and Y points, as shown in Fig. 4(d). These discussions show the long edge state in $Cu_2C_2N_4$ sheet is robust against SOC.

**4.3 Catalytic properties of $Cu_2C_2N_4$ sheet**

The appearance of the long Fermi arc for the open nodal line in $Cu_2C_2N_4$ sheet motivates us to evaluate its electrocatalytic HER activity on the edge, as conceptually displayed in Fig. 5(a). The HER mechanism can be summarized as the following three steps. The first step is the Volmer reaction, during which an electron transfers to a proton and forms an H atom adsorbing on the catalyst surface, described as:

$$H^+ + e^- + * \rightarrow H^* \quad (1)$$

where * and H* represent the active center and the intermediate, respectively. Then,

the H$_2$ desorption realizes following the Tafel reaction or the Heyrovsky reaction, where can be respectively described as:

$$2H^* \rightarrow H_2 + 2^* \quad (2)$$

$$H^+ + e^- + H^* \rightarrow H_2 + ^* \quad (3)$$

where H* still serves as the intermediate. Thus, the HER rate is highly relevant to the binding condition between the intermediate and the active site. As the result, $\Delta G_{H^*}$ of hydrogen adsorption is a crucial parameter to characterize the HER activity. The parameter $\Delta G_{H^*}$ can be obtained by using the formula [55]:

$$\Delta G_{H^*} = \Delta E_H + \Delta E_{ZPE} - T\Delta S_H \quad (4)$$

where $\Delta E_H$ is the adsorption energy for H, $\Delta E_{ZPE}$ and $\Delta S_H$ are the changes in zero-point energy and entropy between the absorbed H and gaseous H, respectively.

For the H adsorption process, here the model is constructed by adsorbing one H atom on edge of the 1×3 supercell of Cu$_2$C$_2$N$_4$ sheet. To be noted, the value of $\Delta G_{H^*}$ is almost unchanged if larger supercell is assigned. By display H atom on the top of Cu, C, and N. We find that the Cu site is the most stable site, while adsorption on other sites was less stable. At the most favorable adsorption, we have calculated the charge density difference to trace the charge transfer during the period. As shown in Fig. 5(b), it is clear that charge depletion occurs on the H atom (see the top panel) while charge accumulation occurs around Cu atoms (see the bottom panel).

Remarkably, under this state, the calculated $\Delta G_{H^*}$ for HER in Cu$_2$C$_2$N$_4$ sheet is as low as 0.10 eV. As compared in Fig. 5(c), this value is greatly lower than those in typical 3D Weyl TQCs including NbP, TaAs, and NbAs (0.31-0.96 eV) [18]. In addition, to compared the performance in Cu$_2$C$_2$N$_4$ sheet with other typical HER catalysts, the volcano curve has been displayed in Fig. 5(d). Notably, $\Delta G_{H^*}$ for HER in Cu$_2$C$_2$N$_4$ sheet almost situates at the top of the volcano curve, indicating the edge of Cu$_2$C$_2$N$_4$ sheet is significantly active, due to its long edge states from the nontrivial nodal line. To be noted, such long edge states still retain after H adsorption except for of a shift of its position [see Fig. 5(b)]. From the volcano curve in Fig. 5(d), we can find that except Pt, $\Delta G_{H^*}$ of Cu$_2$C$_2$N$_4$ sheet is superior to all the HER catalysts displayed in the volcano curve, including the transition metals (Pd, Rh, Ir, Ni, Cu, Ag,

Au, and etc.) and TQCs proposed previously (PtAl, PtGa, NbP, TaAs, NbAs, and etc.) [18-24,56-58]. Meanwhile, the value of $|\Delta G_{H^*}|$ of $Cu_2C_2N_4$ is even comparable with that of Pt (0.10 eV versus 0.09 eV). Thus, $Cu_2C_2N_4$ sheet may serve as a high performance HER catalyst without containing noble metals.

To further capture the relationship of the topological nodal line and the HER activity, we shift the position of nodal line by artificially tuning the number of electrons (Nelec.) in $Cu_2C_2N_4$ sheet. As shown in Fig. 6(a), the nodal line in the S-Y path will be lifted away from the Fermi level when takes out 0.5 e$^-$ from $Cu_2C_2N_4$ sheet. Differently, by adding 0.5 e$^-$ into $Cu_2C_2N_4$ system, the nodal line will be pulled below the Fermi level [see Fig. 6(c)]. The case for native $Cu_2C_2N_4$ sheet is provide in Fig. 6(b) for comparison, where most region of nodal line locates around the Fermi level. Compared with native $Cu_2C_2N_4$ sheet, the nodal line and corresponding surface Fermi arcs in the cases with Nelec.= 49.5 and 50.5 are less contributive to the conducting carriers and the HER process. We further calculate $\Delta G_{H^*}$ for the three cases in Fig. 6(a)-(c). Just as expected, with pulling nodal line away from the Fermi level, $\Delta G_{H^*}$ in $Cu_2C_2N_4$ sheet is significantly increased (8.65 eV for Nelec.= 49.5, and 1.65 eV for Nelec.= 50.5), as shown in Fig. 6(d). These results have fully evidenced the HER enhancement from nodal line and corresponding surface Fermi arcs near the Fermi level in $Cu_2C_2N_4$ sheet.

## 4. Summary

In summary, we have demonstrated the feasibility of developing TQCs for HER in 2D scale. We build an effective design scheme for developing 2D TQCs by taking into account various topological states and corresponding edge features. The design scheme suggests 2D open nodal line with long Fermi arc that transverses the entire edge boundary can maximize the HER activity in 2D TQCs. Following this design scheme, we identify a new 2D material namely $Cu_2C_2N_4$ sheet with open nodal line at the BZ boundary, and theoretically verify that on its edge, where the long Fermi arc exist, the $\Delta G_{H^*}$ for HER is relatively low (0.10 eV), suggesting a high catalytic performance. Especially, the $\Delta G_{H^*}$ in $Cu_2C_2N_4$ sheet nearly situates at the top of the volcano curve, and is comparable with that in Pt. The current work provides a platform to develop TQCs in 2D scale, and paves a feasible way to exploit HER catalysts without containing noble metals as well.

## Acknowledgements

This work is supported by National Natural Science Foundation of China (Grants No. 11904074). The work is funded by Science and Technology Project of Hebei Education Department, the Nature Science Foundation of Hebei Province, S&T Program of Hebei (A2019202107), the Overseas Scientists Sponsorship Program by Hebei Province (C20200319 and C20210330). The work is also supported the State Key Laboratory of Reliability and Intelligence of Electrical Equipment (No. EERI_OY2020001), Hebei University of Technology. One of the authors (X.M. Zhang) acknowledges the financial support from Young Elite Scientists Sponsorship Program by Tianjin.

# Figures and captions:

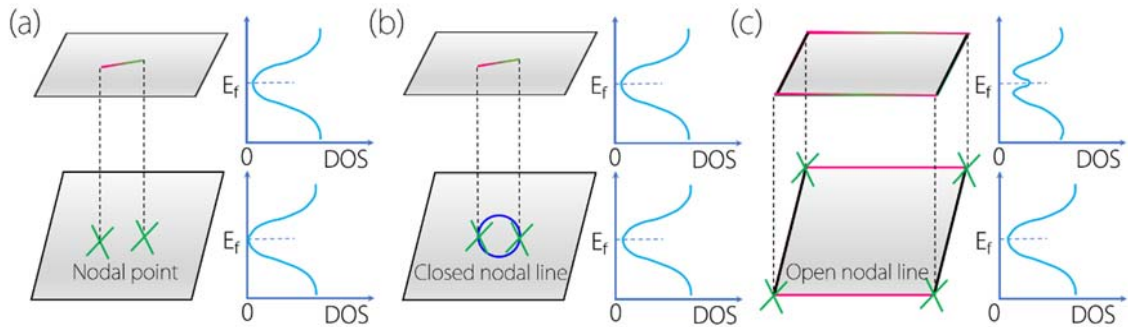

Fig. 1 Design scheme with showing momentum space diagrams and DOSs for 2D Weyl points, 2D closed nodal line, and 2D open nodal line. TWs and DNLs. (a) A pair of Weyl points in 2D plane (left lower panel) and partial Fermi arc (left upper panel) on the edge. Their corresponding DOSs are displayed in right panels. (b) A closed nodal line in the plane (left lower panel) and the partial Fermi arc (left upper panel) on the edge. Their corresponding DOSs are displayed in right panels. (c) A open nodal line at the 2D BZ boundary (left lower panel) and the traversing Fermi arc (left upper panel) on the edge. Their corresponding DOSs are displayed in right panels.

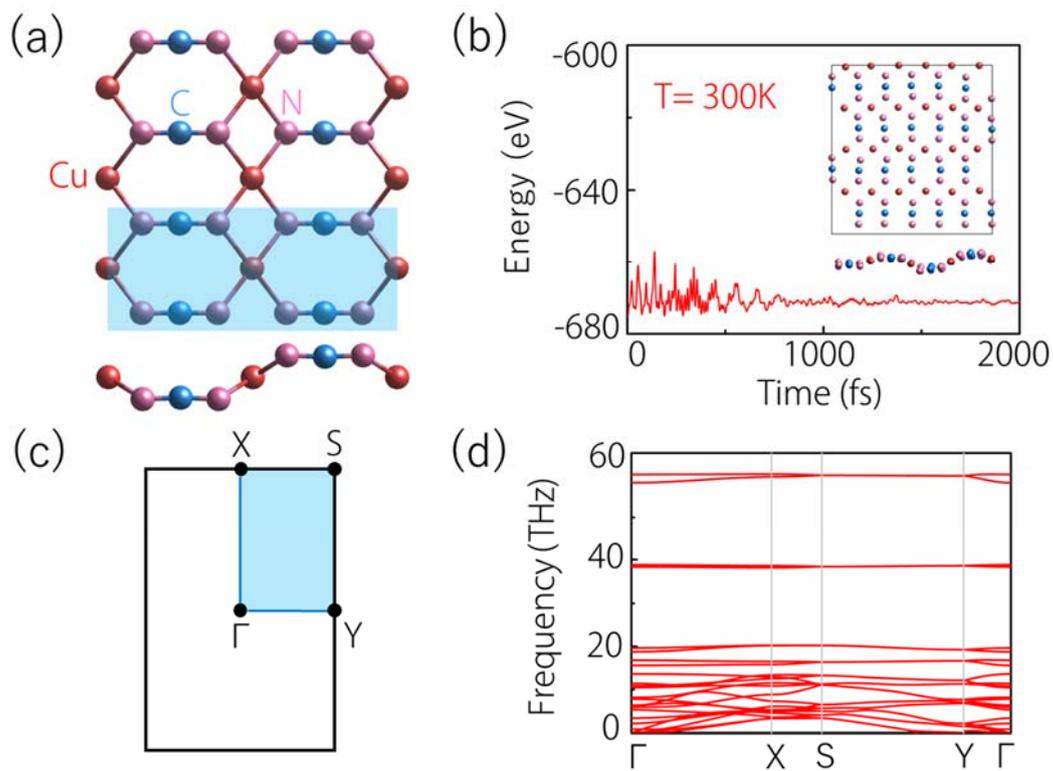

Fig. 2 (a) Crystal structure for $Cu_2C_2N_4$ sheet shown as the top and side views. The shadowed region indicates the primitive cell form of $Cu_2C_2N_4$ sheet. (b) Total potential energy fluctuation of $Cu_2C_2N_4$ sheet during the AIMD simulation. The temperature is set at 300 K. The final state for $Cu_2C_2N_4$ sheet structure after the AIMD simulation is shown as the inset of (b). (c) The BZ of 2D $Cu_2C_2N_4$ sheet. (d) The phonon spectrum for $Cu_2C_2N_4$ sheet.

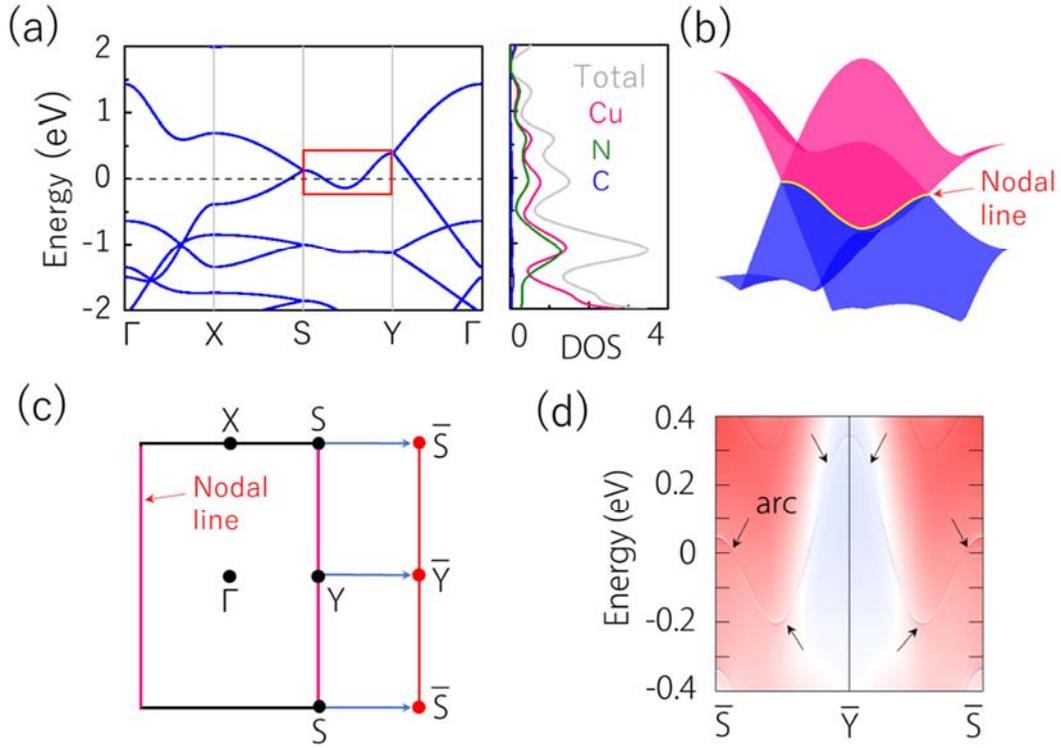

Fig. 3 (a) Electronic band structure, total DOSs and partial DOSs of $Cu_2C_2N_4$ sheet. The framed region in the band structure show the band overlap in the *k*-path S-Y. (b) The 3D plotting of the two bands near the Fermi level. The band crossing (nodal line) is indicated in the figure. (c) The Brillouin zone of the $Cu_2C_2N_4$ sheet and its edge projection. The position of nodal line is indicated in the figure. (d) The projected edge state for the nodal line, as pointed by the arrows.

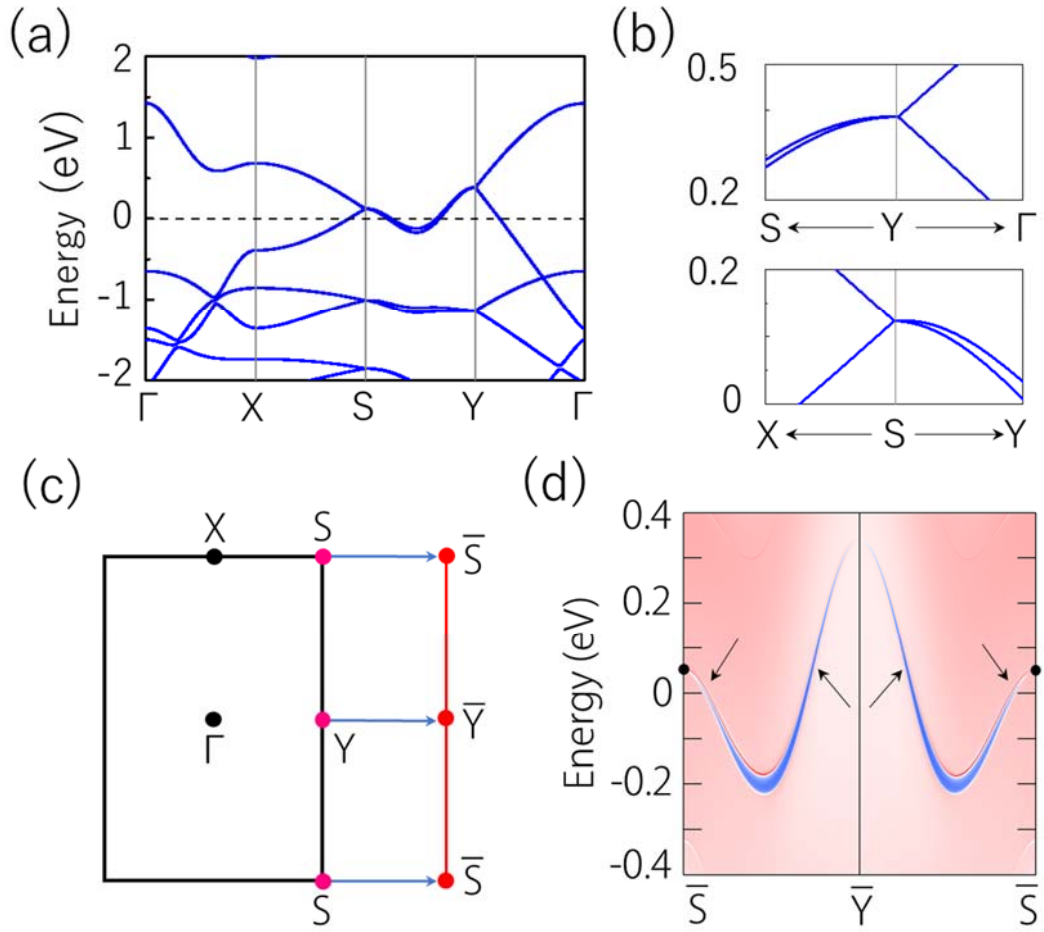

Fig. 4 (a) Electronic band structure for Cu$_2$C$_2$N$_4$ sheet under SOC. (b) Enlarged band structures along the S-Y-Γ and X-S-Y *k*-paths. (c) The position of the crossing point (at the S and Y points) and its edge projection. (c) The projected edge state of Cu$_2$C$_2$N$_4$ sheet, as pointed by the arrows.

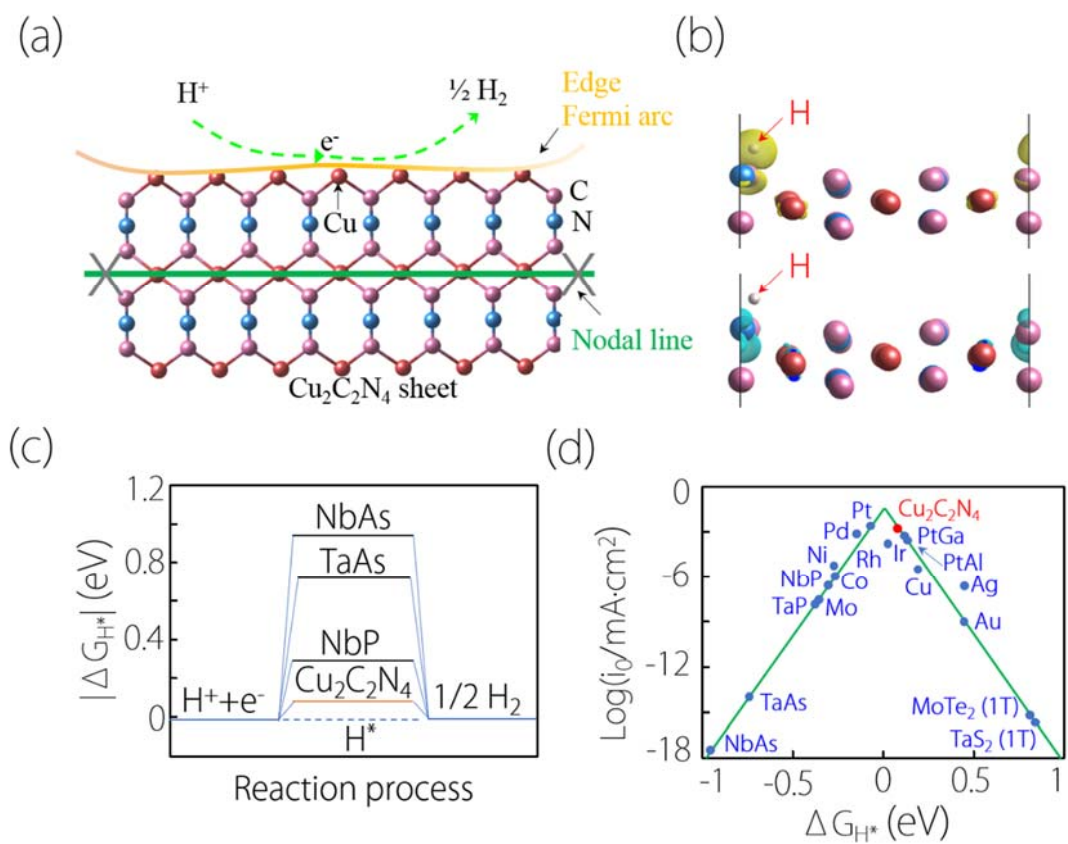

Fig. 5 (a) Illustration of the effects of nodal line and edge Fermi arc on the HER activity in $Cu_2C_2N_4$ sheet. (b) The electron depletion (the top panel) and accumulation (the bottom panel) during H adsorption on edge of $Cu_2C_2N_4$ sheet. The isosurface value is set as 0.008 e Å$^{-3}$. (c) The free energy diagram for HER, where a potential $U$ = 0 is set relative to the standard hydrogen electrode at pH = 0. The free energy of H$^+$ + e$^-$ is by definition the same as that of 1/2 H$_2$ at standard condition of equilibrium. The data of NbAs, NbP (−0.31 eV) and TaAs are picked from Ref. [18]. (d) Volcano plot for the HER of $Cu_2C_2N_4$ sheet in comparison with various pure metals and 3D TQCs. The data for pure metals and 3D TQCs are taken from literatures 18-24,56-58.

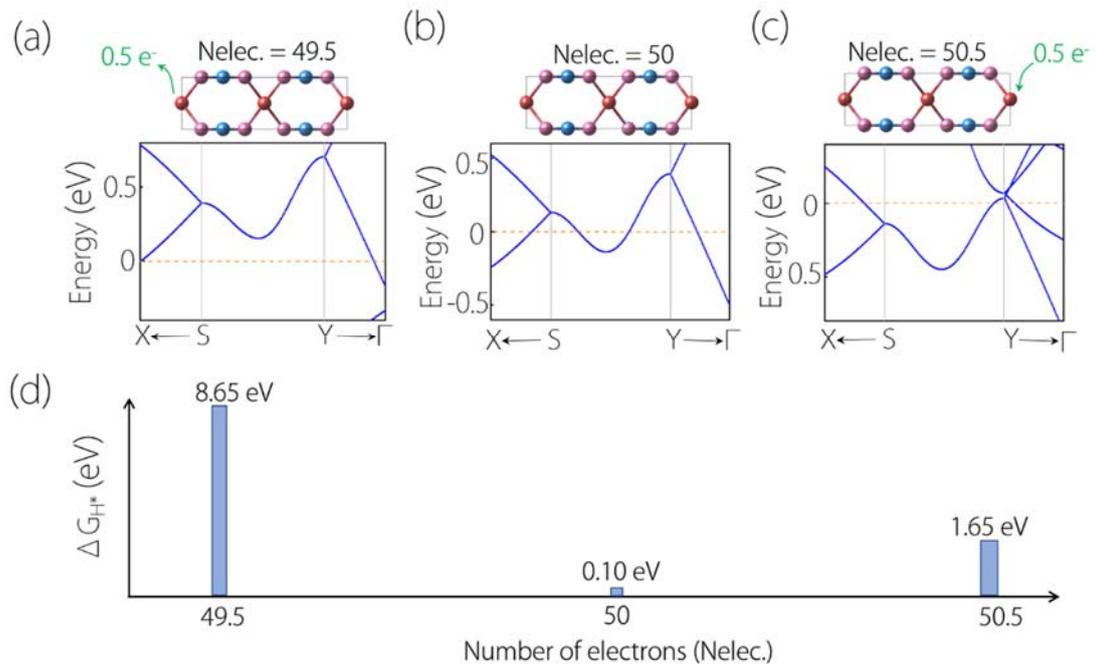

Fig. 6 (a) The model and electronic band structure with taking out 0.5 e⁻ from the Cu$_2$C$_2$N$_4$ system. The native number of electrons (Nelec.) in a Cu$_2$C$_2$N$_4$ cell is 50. (b) and (c) are similar with (a) but for the cases of native Cu$_2$C$_2$N$_4$ and that with adding 0.5 e⁻ into Cu$_2$C$_2$N$_4$ system, respectively. (d) Comparison of ΔG$_{H*}$ in Cu$_2$C$_2$N$_4$ sheet for the cases of Nelec.= 49.5, 50, and 50.5.